\documentclass[pre,showpacs,showkeys,preprintnumbers,amsmath,amssymb,superscriptaddress,twocolumn]{revtex4}

\usepackage{epsfig}
\usepackage{epstopdf}
\usepackage{color}
\usepackage[english]{babel}

\newcommand{\be}{\begin{equation}}
\newcommand{\ee}{\end{equation}}

\newcommand\disp{\displaystyle}

\def\erf{\textrm{erf}}
\def\arcsin{\textrm{arcsin}}
\def\Ei{\textrm{Ei}}

\begin{document}

\title{Overlap of two Brownian trajectories: exact results for scaling functions}

\author{M. V. Tamm}
\affiliation{Physics Department, Moscow State University, 119992, Moscow, Russia}
\affiliation{Department of Applied Mathematics, MIEM, Higher School of Economy, 110000, Moscow, Russia}

\author{V. I. Stadnichuk}
\affiliation{Physics Department, Moscow State University, 119992, Moscow, Russia}

\author{A. M. Ilyina}
\affiliation{Physics Department, Moscow State University, 119992, Moscow, Russia}

\author{D. S. Grebenkov}
\affiliation{Laboratoire de Physique de la Mati\`ere Condens\'ee, CNRS -- Ecole Polytechnique, F-91128 Palaiseau, France}

\date{\today}

\begin{abstract}
We consider two random walkers starting at the same time $t=0$ from
different points in space separated by a given distance $R$.  We
compute the average volume of the space visited by both walkers up to
time $t$ as a function of $R$ and $t$ and dimensionality of space $d$.
For $d<4$, this volume, after proper renormalization, is shown to be
expressed through a scaling function of a single variable
$R/\sqrt{t}$.  We provide general integral formulas for scaling
functions for arbitrary dimensionality $d<4$.  In contrast, we show
that no scaling function exists for higher dimensionalities $d \geq
4$.
\end{abstract}



\maketitle


Statistical properties of the random walk trajectories have been
intensively studied for decades.  The average volume $W_1(t)$ visited
by a single $t$-step Brownian random walk on a $d$-dimensional lattice
was calculated in the 1960s \cite{Vineyard,MW}.  This calculation has
became a part of extended courses of random walk theory (see, for
example \cite{Hughes,Weiss,KRB}), and found many applications in
reaction-diffusion processes \cite{Ben-Avraham} and polymer sciences
\cite{deGennes,Doi}.  A generalization of this classical result to the
case of several random walks is of fundamental interest.  In the 1990s
Larralde and coworkers provided a part of this generalization
\cite{Larralde,Havlin,LWS}.  Namely, they calculated the mean number
of sites visited by {\it at least one} of $N$ walkers starting from
the common origin.  Recently, a complementary question was addressed:
what is the average number of sites $W_N(t)$ visited by {\it all} $N$
walkers \cite{MaTa}.  This quantity as a function of the space
dimensionality $d$ and the number of walkers $N$ was calculated and
its asymptotic behavior for large $t$ was studied.  These results were
rederived using the notion of fractal intersections in \cite{turban}
and further generalized in \cite{MaPRL} where the whole distribution
of the number of sites visited by $N$ walkers was calculated exactly
for $d=1$.

\begin{figure}
\epsfig{file=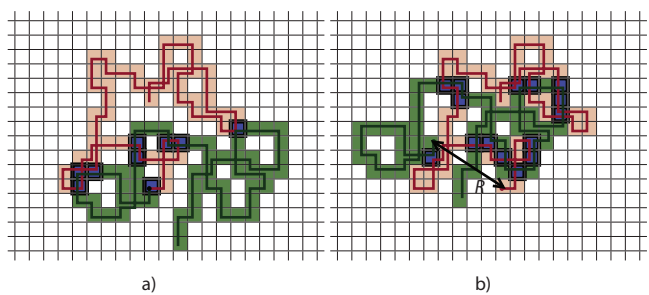, width=8.5 cm} \caption{ (Color online) A realization of two random walks, starting
at the origin (a), and at two points separated by distance $R$ (b).  The sites visited by \emph{both} walks
are denoted by dark squares.}
\label{fig_common}
\end{figure}

In this paper, we propose a different generalization of \cite{MaTa}.
We consider random walks that, instead of starting altogether from the
origin $x=0$, have {\it distinct} starting points $x_i$ ($i=1\ldots
N$).  The values of $x_i$ (or, more precisely, $x_i-x_j$) will
influence the behavior of $W_N(t)$, which now is denoted as $W_N(t,
{x_i})$.  For large enough $t$, however, random walks ``forget'' their
initial positions, and the position-dependent function $W_N(t, {x_i})$
should converge to $W_N(t)$ studied in
\cite{MaTa}.  More generally, one can write
\begin{equation}
W_N(t,{x_i}) = W_N(t)~  \Phi_{d}(\xi_i,\ldots,\xi_{N-1}),
\label{intro1}
\end{equation}
where $\Phi_{d}(\xi_i,\ldots,\xi_{N-1})$ is a function of scaling variables $\xi_i \sim (x_i - N^{-1}
\sum_{j=1}^N x_j)/\sqrt{t}$ (exact prefactor will be chosen below)
\footnote{ Note that there are only $N-1$ independent scaling variables.  It is convenient to consider the
starting positions of the walkers relative to their center of mass.},
and $d$ is the dimensionality of space.  This scaling function should
converge to unity as $\xi_i \to 0$ ($i=1\ldots N-1$), and to zero if
at least one of $x_i$ is much larger than $1$ (indeed, if the starting
positions are separated by a distance much larger than $t^2$, the
probability of any overlap is exponentially small).  In this paper, we
show that the scaling function $\Phi_{d}(\xi)$ can be calculated
exactly in the case of two random walkers starting at a distance $R$
from each other (see Fig. \ref{fig_common}).

We consider two random walks of given lengths $t_1$, $t_2$, their
starting points $x_1$, $x_2$ being separated by the distance
$R=|x_1-x_2|$.  We are interested in calculating the average volume
$w_2$ of the domain visited by both random walkers as a function of
$t_1$, $t_2$ and $R$.  Following \cite{MaTa} we express this volume in
terms of the probability that a given site $x$ is visited by each of
the walkers
\begin{equation}
w_2 (x_1,x_2,t_1,t_2) = \int p(x,t_1|x_1) p(x,t_2|x_2)d^d x.
\label{eq1}
\end{equation}
Here $p(x,t|x_0)$ is the probability that a random walk starting at
$x_0$ has visited a point $x$ by time $t$, and the integral should be
replaced with the sum for walkers in discrete space (lattice).

Now, the probability $p(x,t|x_0)$ can be expressed in terms of the
random walk propagator $g(x,t|x_0)$ as follows
\begin{equation}
p(x,t|x_0)=\int_0^t g(x,\tau|x_0) q(t-\tau|x) d \tau ,
\label{eq2}
\end{equation}
where $q(t|x)$ is the persistence probability at point $x$, i.e. the
probability that a walker starting from the point $x$ does not return
to it up to time $t$.  There is a simple relation between the
persistence probability and the probability $f(t|x)$ of the first
return at the point $x$:
\begin{equation}
\frac{\partial q(t|x)}{\partial t}=-f(t|x), \qquad q(0|x) = 1,
\end{equation}
from which
\begin{equation}
q(t|x) = 1- \int_0^t f(t'|x) dt' .
\end{equation}
Substituting Eq. (\ref{eq2}) into Eq. (\ref{eq1}) yields
\begin{equation}
\begin{split}
w_2 & (x_1,x_2,t_1,t_2) = \int\limits_0^{t_1} d\tau_1\int\limits_0^{t_2} d\tau_2 \int q(t_1-\tau_1|x) \\
& \times  q(t_2-\tau_2|x) g(x,\tau_1|x_1) g(x,\tau_2|x_2)d^d x . \\
\end{split}
\label{eq3}
\end{equation}
In the most interesting case of time-reversible translationally
invariant random walks, $q(t|x) = q(t)$ is site-independent, and
$g(x,t|x_0)=g(x_0,t|x)$, so that Eq. (\ref{eq3}) can be further
simplified into
\begin{equation}
\begin{split}
& w_2 (x_1,x_2,t_1,t_2) \\
& = \int\limits_0^{t_1} d\tau_1 \int\limits_0^{t_2} d\tau_2  ~q(t_1-\tau_1) ~q(t_2-\tau_2)~ g(x_2,\tau_1 + \tau_2|x_1) . \\
\end{split}
\label{eqmaster}
\end{equation}

For Brownian random walks in the whole space $\mathbb{R}^n$, the
propagator is Gaussian (for discrete space, it is asymptotically
Gaussian at large $t$)
\begin{equation}
g(x_1,t|x_2) = (4 \pi Dt)^{-d/2} \exp\left(- \frac{|x_1-x_2|^2}{4Dt}\right),
\label{Gaussian}
\end{equation}
%
where the diffusion coefficient $D = a^2/(2d\delta)$ is related to a
microscopic length $a$ of the order of underlying lattice spacing, and
$\delta$ being the duration of time step.  In addition, the
persistence was well-studied and its asymptotic behavior depends
crucially on the dimensionality of space (in particular, on whether
the walk is recurrent or transient) \cite{Hughes,Bray13}:
\begin{equation}
q(t) \sim \begin{cases} \displaystyle  t^{-d/2} \hskip 27mm (d<2), \cr \displaystyle  (\ln t)^{-1} \hskip 24mm (d=2),
\cr \displaystyle  const + O(t^{-(d-2)/2}) \quad (d>2),  \end{cases}
\label{persistence}
\end{equation}
where the proportionality constants depend in general on both $d$ and
the structure of the underlying lattice.  In what follows, we
substitute Eqs. (\ref{Gaussian}, \ref{persistence}) into Eq.
(\ref{eqmaster}) for the particular case $t=t_1=t_2$ in order to
calculate the scaling function
\begin{equation}
\Phi_d (\xi) \equiv \frac{w_2 (0,R,t,t)}{w_2 (0,0,t,t)},  \qquad \xi \equiv \frac{R}{\sqrt{4dDt}} = \frac{R/a}{\sqrt{2t/\delta}} .
\label{scaling}
\end{equation}
It is convenient to consider separately the two cases (i) $d<2$, and
(ii) $d\geq 2$.

(i) \emph{Dimensionality less than 2.}
Substituting Eqs. (\ref{eqmaster}, \ref{Gaussian}, \ref{persistence})
into Eq. (\ref{scaling}) yields the scaling function in the following
dimensionless form
\begin{equation}
\begin{split}
\Phi_d(\xi) & \simeq  \frac{I_< (\xi,d)}{I_< (0,d)} , \\
 I_<(\xi,d)    & =  \disp \int_0^1 \int_0^1
\frac{dz_1}{(1-z_1)^{d/2}}\frac{dz_2}{(1-z_2)^{d/2}}\frac{\exp(-\frac{d\xi^2}{z_1+z_2})}{(z_1+z_2)^{d/2}} .\\
\end{split}
\label{scaling1}
\end{equation}
This integral can be further simplified by changing variables as
$u=(z_1+z_2)$, $v =(z_1-z_2)/2$:
%
\begin{equation}
\begin{split}
I_<(\xi,d) &=  2 \int_0^{1} du  \frac{(1-u/2)^{1-d}}{u^{d/2}} \exp (-d\xi^2/u) \\
 & \times \int_0^{u/(2-u)} \frac {dv} {(1-v^2)^{d/2}} \\
 & + \frac{\sqrt{\pi}~ \Gamma(1-d/2)}{\Gamma(3/2-d/2)}  \\
 &\times \int_1^2 du~ u^{-d/2} (1-u/2)^{1-d} \exp (-d\xi^2/u).  \\
\end{split}
\label{integralsmall}
\end{equation}
In particular, for $d=1$ one can further simplify this formula into
\begin{equation}
\begin{split}
I_<(\xi,1) &=  2 \int_0^{1} \frac{du}{\sqrt{u}} \exp (-\xi^2/u) \arcsin(u/(2-u)) \\
&+ 2\pi \bigl[\sqrt{2} \exp(-\xi^2/2)-\exp(-\xi^2)\bigr]  \\
& +2 \pi \sqrt{\pi}~ \xi \bigl[\erf(\xi/\sqrt{2}) - \erf(\xi)\bigr] ,
\end{split}
\label{integral1}
\end{equation}
where
\begin{equation*}
\erf(x) = \frac{2}{\sqrt{\pi}} \int_0^x \exp (-y^2) dy
\end{equation*}
is the error function.

Expanding the above expression into a series in the vicinity of
$\xi=0$ leads to
\begin{equation}
\Phi_1(\xi)\simeq 1 - \xi^2 \frac{1}{2(\sqrt{2}-1)} + O(\xi^4) . 
\label{smallxi_1D}
\end{equation}
In turn, for large $\xi$, the error function converges to one
exponentially fast, thus the whole expression in Eq. (\ref{integral1})
vanishes exponentially fast.

(ii) \emph{Dimensionality greater than or equal to two.}  For $d \geq
2$ the scaling function gets an even simpler form.  Indeed, one easily
sees substituting Eqs. (\ref{eqmaster}, \ref{Gaussian},
\ref{persistence}) into Eq.  (\ref{scaling}) that in the first
approximation the input from the persistence cancels out and the
scaling function reads simply
\begin{equation}
\begin{split}
\Phi_d(\xi) & \simeq  \frac{I_> (\xi,d)}{I_> (0,d)}, \\
 I_>(\xi,d)    & =  \int\limits_0^1 \int\limits_0^1 dz_1 dz_2\frac{\exp(-\frac{d\xi^2}{z_1+z_2})}{(z_1+z_2)^{d/2}}. \\
\end{split}
\label{scalinglarge}
\end{equation}
However, for $d \geq 4$ the integral $I_>(0,d)$ diverges which means (see \cite{MaTa}) that the overlap in
this case is controlled by the behavior at small $t$, and scaling function does not exist.  For $d=2,3$ the
integrals in Eq. (\ref{scalinglarge}) can be computed exactly by substitution $u=z_1+z_2$, $v=(z_1-z_2)/2$:
\begin{equation}
\begin{split}
 I_> (\xi,2) & = 2 \bigl[\exp(-2x^2)-\exp(-x^2)\bigr] + \\
 & + 2 (1+2x^2) \Ei (-2x^2) - 2 (x^2+1) \Ei(-x^2), \\
\end{split}
\label{integral2}
\end{equation}
where
\begin{equation*}
\Ei(x)= -\int_{-x}^\infty \exp(-y)/y ~dy
\label{ei}
\end{equation*}
is the exponential integral function; and
\begin{equation}
\begin{split}
& I_>(\xi,3)
 = 4 \exp(-3 \xi^2) -2 \sqrt{2} \exp (-3 \xi^2/2)- 2\sqrt{3\pi}~ \xi \\
& + \sqrt{3\pi} \biggl(4 \xi +\frac{2}{3\xi}\biggr) \erf(\sqrt{3} \xi) -
\sqrt{3\pi} \biggl(2 \xi + \frac{2}{3 \xi}\biggr) \erf(\sqrt{3/2}\xi) . \\
\end{split}
\label{integral3}
\end{equation}
For small $\xi$, the scaling functions behave as
\begin{equation}
\begin{split}
\Phi_2(\xi) &\simeq 1 + \frac {2 \ln \xi}{\ln2} \xi^2 + \frac{2 \ln 2 +\gamma - 2}{\ln 2} \xi^2 - 3 \xi^4/2
 + O (\xi^6), \\
 \end{split}
\label{smallxi_2D}
\end{equation}
where $\gamma \approx 0.577216 $ is the Euler's gamma constant, and
\begin{equation}
\Phi_3(\xi) \simeq 1- \xi \frac{\sqrt{3\pi}}{2\sqrt{2}(\sqrt{2}-1)}+\xi^2 \frac{3+\sqrt{2}}{2} + O (\xi^4)  .
\label{smallxi_3D}
\end{equation}
Note that contrary to Eq. (\ref{smallxi_1D}), the scaling function in $d=2,3$ has a singularity in the
vicinity of $R=0$: indeed, the first corrections are proportional to $R^2 \ln R$ and
$R=\sqrt{R_x^2+R_y^2+R_z^2}$, respectively. For $\xi \gg 1$, both $I_>(\xi,2)$ and $I_>(\xi,3)$ vanish
exponentially fast as expected.

%
%
%

\begin{figure}
\epsfig{file=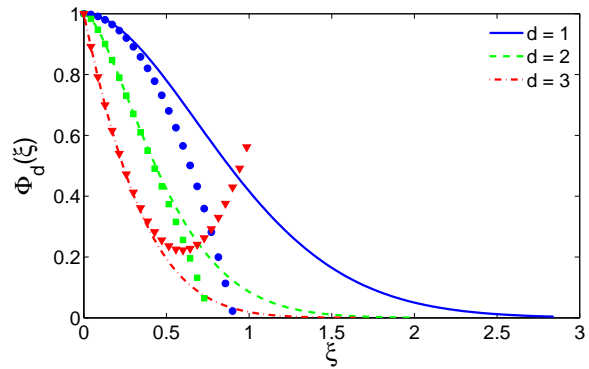, width=8.5cm}
\caption{(Color online) Scaling functions $\Phi_d(\xi)$ for
$d=1,2,3$ (lines) and their asymptotic behaviors in
Eqs. (\ref{smallxi_1D}, \ref{smallxi_2D}, \ref{smallxi_3D}) for
$0<\xi<1$ (symbols).  }
\label{fig_scalingfunctions}
\end{figure}

\begin{figure}
\epsfig{file=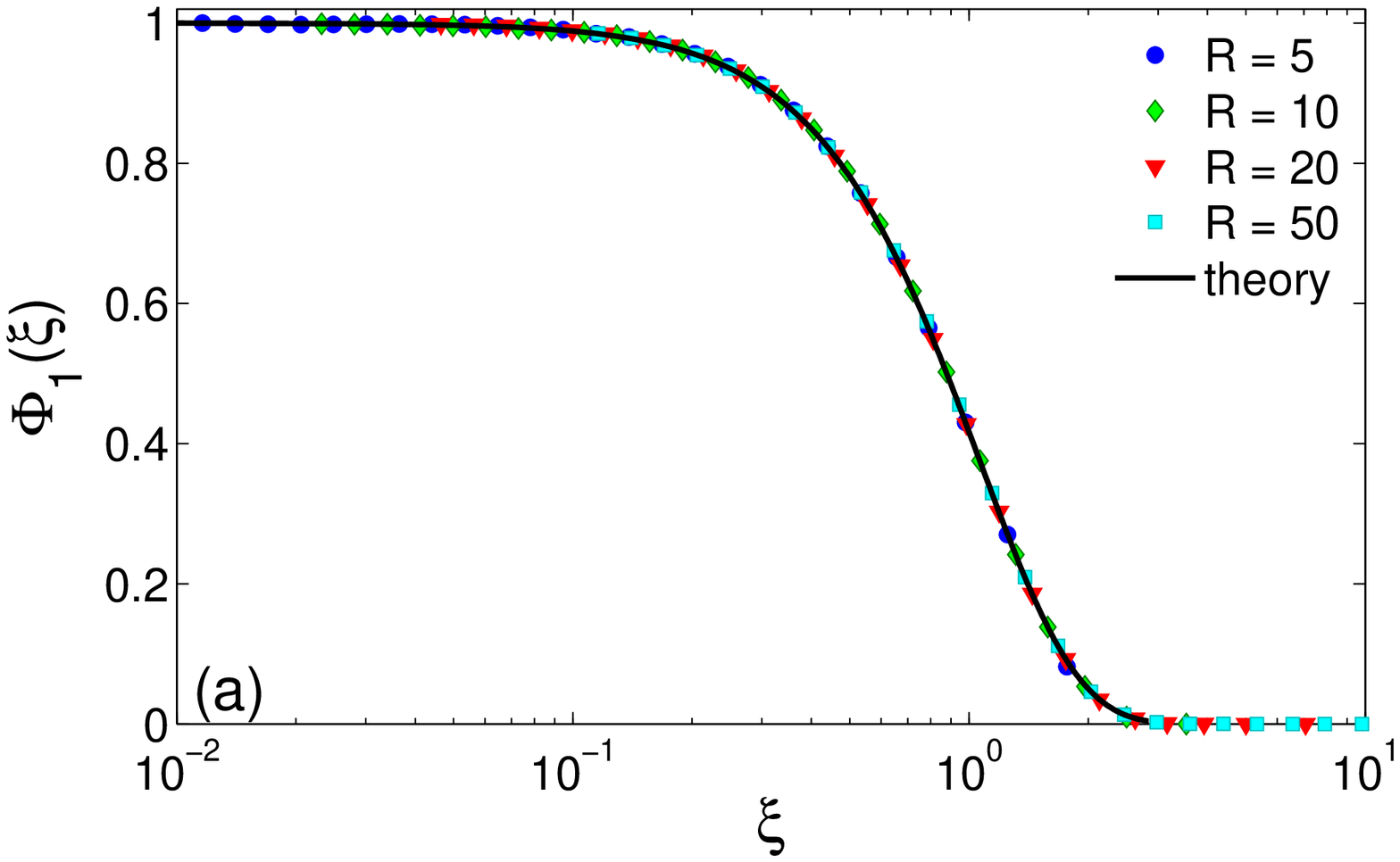, width=8cm}
\epsfig{file=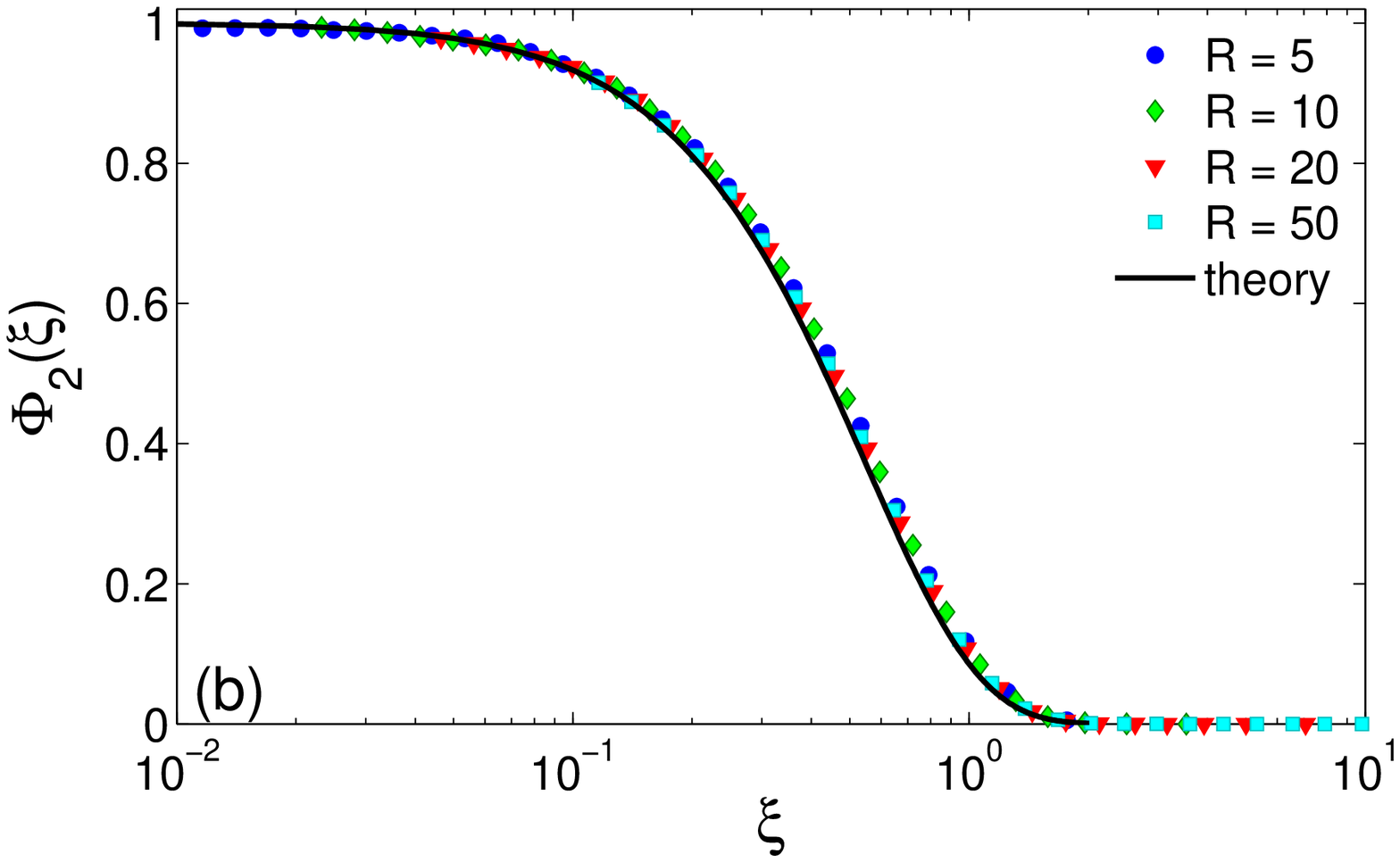, width=8cm}
\epsfig{file=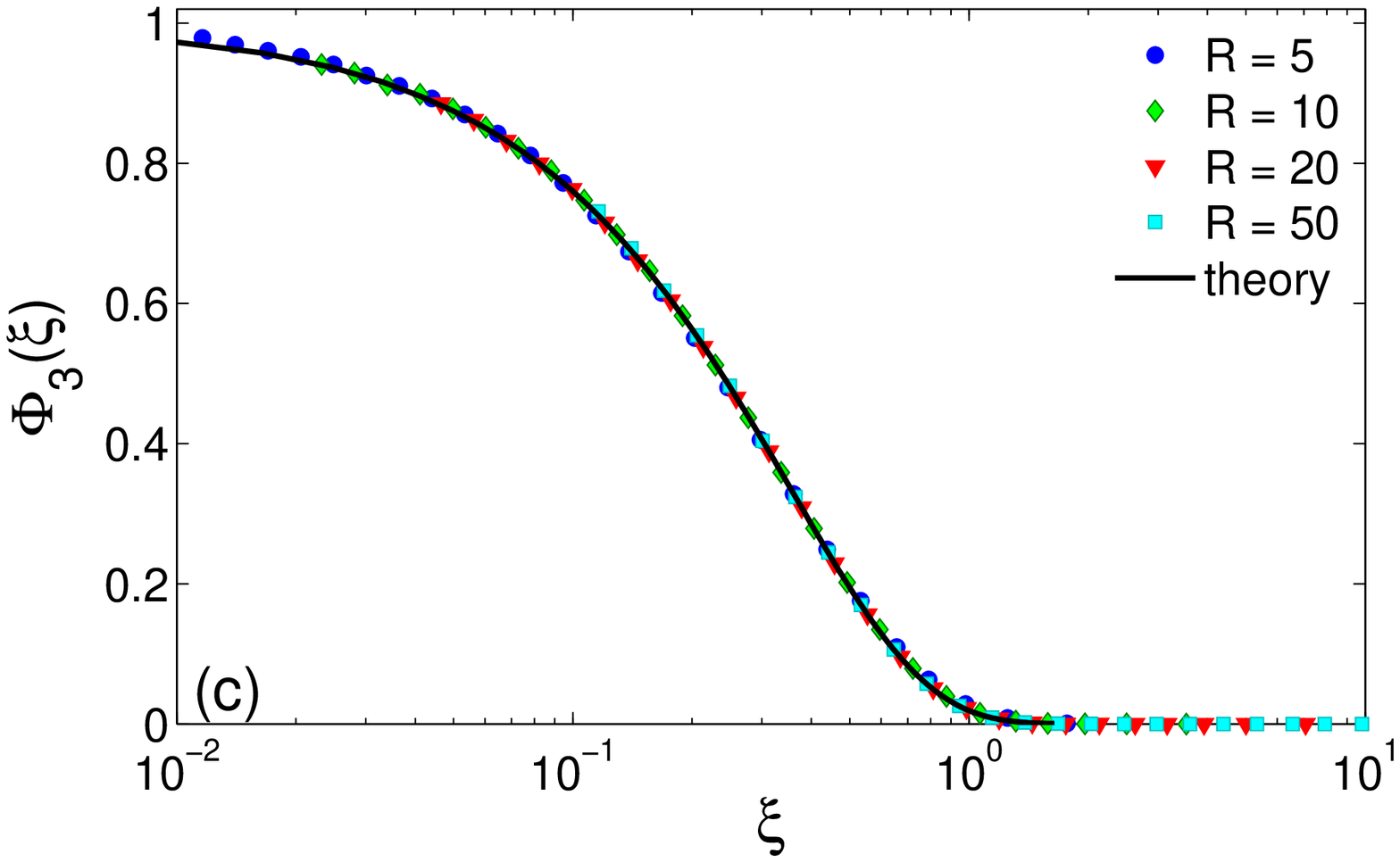, width=8cm}
\epsfig{file=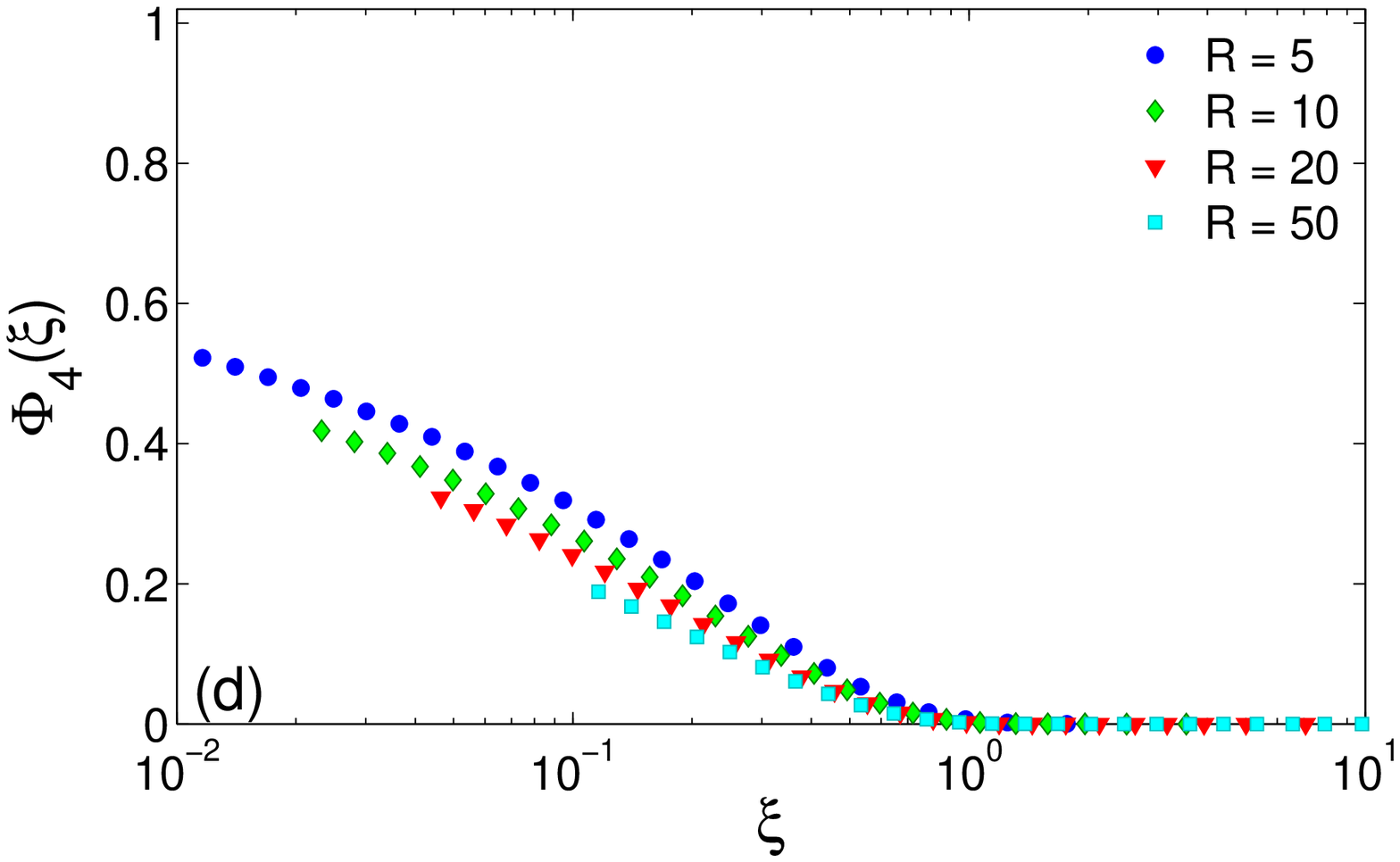, width=8cm}
\caption{ (Color online).  Numerical results for the
renormalized overlap functions $\Phi_d(\xi) =
w_2(0,R,t,t)/w_2(0,0,t,t)$ for $d=1$ (a), $d=2$ (b), $d=3$ (c), and
$d=4$ (d).  Numerical results were obtained by Monte Carlo simulations
of $d$-dimensional random walks on (hyper)cubic lattices with $R$ = 5
(circles), 10 (diamonds), 20 (triangles), and 50 (stars).  Each point
is an average over 262 144 realizations of random walks up to $t=$ 131
072 steps.  The theoretical predictions for the scaling curves in
$d=1,2,3$ are shown with solid black lines.}
\label{fig_numerics}
\end{figure}

Figure \ref{fig_scalingfunctions} shows the scaling functions
$\Phi_d(\xi)$ for $d=1,2,3$ and their asymptotic behaviors.  To check
the results presented above, we simulated random walks on a
(hyper)cubic lattice in $d=1,2,3,4$ for initial distances equal to
$R=5,10,20,50$.  The results are presented in Fig. \ref{fig_numerics}
(note the logarithmic scale of the horizontal axis).  The theoretical
results given by Eqs. (\ref{integral1}), (\ref{integral2}), and
(\ref{integral3}) are shown by thick black lines.  Note the absence of
any scaling collapse of the curves for $d=4$.

In summary, we have shown that the average volume visited by
\emph{both} independent random walkers starting initially at some
given distance $R$ from each others behaves in a strikingly universal
way as a function of the scaling variable $\xi = R/\sqrt{2t}$.  It is
instructive to consider the results in terms of three different phases
of the overlap scaling discussed in \cite{MaTa}.  In the
low-dimensionality phase where overlap for $\xi = 0$ scales as
$w_2(0,0,t,t) \sim t^{d/2}$, corrections due to a non-zero initial
distance are analytical functions of that distance
\begin{equation}
\frac{w_2(0,R,t,t)}{w_2(0,0,t,t)} =  1 - a(d) R^2/t + O (R^4/t^2),
\label{corr_smalld}
\end{equation}
with some $d$-dependent correction constant $a(d)$, e.g., $a(1)=
\frac{1}{4(\sqrt{2}-1)}$ according to Eq. (\ref{smallxi_1D}).  In the
medium-dimensionality phase $2<d<4$, where $w_2(0,0,t,t) \sim
t^{(4-d)/2}$, the introduction of a non-zero initial distance gives
rise to a correction which is singular at $R=0$
\begin{equation}
\begin{split}
 \frac {w_2(0,R,t,t)}{w_2(0,0,t,t)} & = 1 - a(d) (R^2/t)^{(4-d)/2} + O (R^2/t),\\
 & \quad \quad (2<d<4), \\
\frac{w_2(0,R,t,t)}{w_2(0,0,t,t)} & = 1 + \frac {\ln (R^2/t)}{\ln2} R^2/t + O (R^2/t), \\
 & \quad \quad (d=2) . \\
\end{split}
\label{corr_larged}
\end{equation}
Finally, in the large-dimensionality phase with $d\geq 4$ where the
overlap is mostly controlled by the small $t$ behavior of the walks,
no scaling function exists at all.

It is also instructive to consider the difference $w_2(0,0,t,t) -
w_2(0,R,t,t)$, i.e. the average ``deficiency'' of the overlap function
due to the walks starting at distance $R$ from each other.  For $d<2$
this difference converges to zero at large $t$ as $R^2 t^{(d-2)/2}$,
while for $2<d<4$ it converges to a finite limit which scales as
$R^{4-d}$.

One other interesting qualitative result illustrating the behavior in
the $2<d<4$ region concerns the fraction of the sites visited by the
first walk which are also visited by the second walk, i.e. the ratio
\begin{equation}
f_d(R,t) = \frac{w_2(0,R,t,t)}{w_1(t)} = \frac{w_2(0,R,t,t)}{w_2(0,0,t,t)}\frac{w_2(0,0,t,t)}{w_1(t)}.
\label{frac}
\end{equation}
The two ratios on the right hand-side of Eq. (\ref{frac}) are both
positive and converge to zero as $t\to 0$ or $t \to \infty$.
Therefore, for any given $R$ there exists time (of order $R^2$) at
which the relative overlap of two walks is maximal:
$f_d(R,t)=f_d^{\max}(R) \sim R^{2-d}$.

The results on the average volume of several random walks can be of practical use to estimate, e.g. the
interactions and entanglements of Gaussian polymer coils, or the oversampling rate in intermittent search
processes where the search for the target is an alternating sequence of random walks and longer jumps (see
\cite{intermit} for examples), or surface-mediated diffusion \cite{OTV,Calandre12,Calandre14}.

In order to keep the presentation as transparent as possible, we
concentrated here on the simplest possible set-up of two walks of
equal length.  It is clear that the generalization for walks of
different lengths is straightforward, and all asymptotical results for
$t \to \infty$ hold as soon as the two walk lengths remain comparable
in this limit.  Generalizations for $N>2$ walkers are more cumbersome
but also straightforward.  It may be also interesting to study the
overlaps further in confined geometries (e.g., in a $d$-dimensional
sphere or torus): in this case the overlap fraction $f_d(R,t)$ can
exhibit a peculiar non-monotonous behavior as a function of $t$.


\begin{acknowledgments}
The authors are grateful to S.N. Majumdar for valuable discussions. This work was partially supported by the
grant FP7-PEOPLE-2010-IRSES 269139 DCP-PhysBio and by ANR project ``INADILIC''.
\end{acknowledgments}


\begin{thebibliography}{99}

\bibitem{Vineyard} G.~H. Vineyard, J. Math. Phys. {\bf 4}, 1191 (1963).

\bibitem{MW} E.~W. Montroll and G.~H. Weiss, J. Math. Phys. {\bf 6}, 167 (1965).

\bibitem{Hughes} B.~H. Hughes, {\em Random Walks and Random Environments, vol. 1}
(Clarendon Press, Oxford, 1996);

\bibitem{Weiss}  G.~H. Weiss, {\em Aspects and Applications of the Random Walk}
(North-Holland, Amsterdam, 1994).

\bibitem{KRB} P. Krapivsky, S. Redner, E. Ben-Naim, {\em A Kinetic View of Statistical Physics} (Cambridge
University Press, 2010).

\bibitem{Ben-Avraham} D. Ben-Avraham and S. Havlin, {\em Diffusion and reaction in disordered systems}
                           (Cambridge University Press, 2000).

\bibitem{deGennes} P. G. de Gennes, {\em Scaling Concepts in Polymer Physics}
                           (Cornell University, Ithaca, New York, 1979).

\bibitem{Doi}  M. Doi and S. F. Edwards,
                           {\em The Theory of Polymer Dynamics}
                           (Oxford University Press, Oxford, UK, 1986).



\bibitem{Larralde} H. Larralde, P. Trunfio, S. Havlin, H.~E. Stanley, and G.~H. Weiss,
Nature (London) {\bf 355}, 423 (1992); Phys. Rev. A {\bf 45}, 7128 (1992).

\bibitem{Havlin} S. Havlin, H. Larralde, P. Trunfio, J.E. Kiefer, H.~E. Stanley, and G.~H.
Weiss, Phys. Rev. A {\bf 46}, R1717 (1992).

\bibitem{LWS} J. Larralde, G.~H. Weiss, and H.~E. Stanley, Physica A {\bf 209}, 361 (1994).

\bibitem{MaTa} S.N. Majumdar, M.V. Tamm, Phys. Rev. E {\bf 86}, 021135 (2012).

\bibitem{MaPRL} A. Kundu, S.N. Majumdar, G. Schehr, Phys. Rev. Lett. {\bf 110}, 220602 (2013).

\bibitem{turban} L. Turban, arXiv:1209.2527.

\bibitem{Bray13} A. J. Bray, S. N. Majumdar, and G. Schehr,
                           Adv. Phys. {\bf 62}, 225-361 (2013).

\bibitem{intermit} O.B\'{e}nichou, C. Lovredo, M. Moreau, and R. Voituriez, Rev. Mod. Phys. {\bf 83}, 81
(2011).

\bibitem{OTV} G. Oshanin, M. Tamm, O. Vasilyev, J. Chem. Phys., {\bf 132}, 235101 (2010).

\bibitem{Calandre12}  T. Calandre, O. B\'enichou, D. S. Grebenkov, and R. Voituriez,
Phys. Rev. E {\bf 85}, 051111 (2012).

\bibitem{Calandre14}  T. Calandre, O. B\'enichou, D. S. Grebenkov, and R. Voituriez,
Phys. Rev. E {\bf 89}, 012149 (2014).

\end{thebibliography}
\end{document}